\documentclass[preprint,preprintnumbers,amsmath,amssymb,superscriptaddress]{revtex4}

\usepackage{graphicx}
\usepackage{dcolumn}
\usepackage{bm}

\begin{document}

\centerline {SCIENCE \textbf{323}, 919 (2009) [Perspectives at
SCIENCE \textbf{323}, 888 (2009)]}


\title{First direct observation of spin textures in topological insulators : Spin-resolved ARPES as a probe of topological quantum spin Hall and Berry's phase effects}

\author{D. Hsieh}
\affiliation{Joseph Henry Laboratories of Physics, Princeton
University, Princeton, NJ 08544, USA}
\author{Y. Xia}
\affiliation{Joseph Henry Laboratories of Physics, Princeton
University, Princeton, NJ 08544, USA}
\author{L. Wray}
\affiliation{Joseph Henry Laboratories of Physics, Princeton
University, Princeton, NJ 08544, USA}
\author{D. Qian}
\affiliation{Joseph Henry Laboratories of Physics, Princeton
University, Princeton, NJ 08544, USA}
\author{A. Pal}
\affiliation{Joseph Henry Laboratories of Physics, Princeton
University, Princeton, NJ 08544, USA}
\author{J. H. Dil}
\affiliation{Swiss Light Source, Paul Scherrer Institute, CH-5232,
Villigen, Switzerland} \affiliation{Physik-Institut, Universit\"{a}t
Z\"{u}rich-Irchel, 8057 Z\"{u}rich, Switzerland}
\author{F. Meier}
\affiliation{Swiss Light Source, Paul Scherrer Institute, CH-5232,
Villigen, Switzerland} \affiliation{Physik-Institut, Universit\"{a}t
Z\"{u}rich-Irchel, 8057 Z\"{u}rich, Switzerland}
\author{J. Osterwalder}
\affiliation{Physik-Institut, Universit\"{a}t Z\"{u}rich-Irchel,
8057 Z\"{u}rich, Switzerland}
\author{G. Bihlmayer}
\affiliation{Institut f\"{u}r Festk\"{o}rperforschung,
Forschungszentrum J\"{u}lich, D-52425 J\"{u}lich, Germany}
\author{C. L. Kane}
\affiliation{Department of Physics and Astronomy, University of
Pennsylvania, Philadelphia, PA 19104, USA}
\author{Y. S. Hor}
\affiliation{Department of Chemistry, Princeton University,
Princeton, NJ 08544, USA}
\author{R. J. Cava}
\affiliation{Department of Chemistry, Princeton University,
Princeton, NJ 08544, USA}
\author{M. Z. Hasan}
\affiliation{Joseph Henry Laboratories of Physics, Princeton
University, Princeton, NJ 08544, USA} \affiliation{Princeton Center
for Complex Materials, Princeton University, Princeton, NJ 08544,
USA}\email{mzhasan@Princeton.edu}

\date{First submitted to Science on July 22$^{nd}$, 2008}

\begin{abstract}
\textbf{A topologically ordered material is characterized by a rare quantum
organization of electrons that evades the conventional spontaneously
broken symmetry based classification of condensed matter. Exotic
spin transport phenomena such as the dissipationless quantum spin
Hall effect have been speculated to originate from a novel
topological order whose identification requires a spin sensitive
measurement, which does not exist to this date in any system (neither in Hg(Cd)Te quantum wells nor in the topological insulator BiSb). Using spin-ARPES in a Mott polarimetric mode, we probe the spin degrees of freedom of these quantum spin Hall states and demonstrate that topological quantum numbers are uniquely determined
from spin texture imaging measurements. Applying this method to the Bi$_{1-x}$Sb$_x$ series, we identify the origin of its novel order and
unusual chiral properties. These results taken together constitute
the first observation of surface electrons collectively carrying a
geometrical quantum Berry's phase and definite chirality (mirror Chern number, $n_M$ =-1), which are the key electronic properties for realizing topological
computing bits with intrinsic spin Hall-like topological phenomena. Our spin-resolved results not only provides the first clear proof of a topological insulating state in nature but also demonstrate the utility of spin-resolved-ARPES technique in measuring the quantum spin Hall phases of matter.}
\end{abstract}

\maketitle

Ordered phases of matter such as a superfluid or a ferromagnet are
usually associated with the breaking of a symmetry and are
characterized by a local order parameter [1], and the typical
experimental probes of these systems are sensitive to order
parameters. The discovery of the quantum Hall effects in the 1980s
revealed a new and rare type of order that is derived from an
organized collective quantum motion of electrons [2-4]. These
so-called ``topologically ordered phases" do not exhibit any
symmetry breaking and are characterized by a topological number [5]
as opposed to a local order parameter. The classic experimental
probe of topological quantum numbers is magneto-transport, where
measurements of the quantization of Hall conductivity $\sigma_{xy} =
ne^{2}/h$ (where $e$ is the electric charge and $h$ is Planck's
constant) reveals the value of the topological number $n$ that
characterizes the quantum Hall effect state [6].

Recent theoretical and experimental studies suggest that a new class
of quantum Hall-like topological phases can exist in spin-orbit
materials without external magnetic fields, with interest centering
on two examples, the ``quantum spin Hall insulator" [7-9] and the
``strong topological insulator" [10,11]. Their topological order is
believed to give rise to unconventional spin physics at the sample
edges or surfaces with potential applications ranging from
dissipationless spin currents [12] to topological (fault-tolerant)
quantum computing [13]. However, unlike conventional quantum Hall
systems, these novel topological phases do not necessarily exhibit a
quantized charge or spin response ($\sigma_{xy} \neq ne^{2}/h$)
[14,15]. In fact, the spin polarization is not a conserved quantity
in a spin-orbit material. Thus, their topological quantum numbers,
the analogues of $n$, cannot be measured via the classic von
Klitzing-type [2] transport methods.

Here we show that spin-resolved angle-resolved photoemission
spectroscopy (spin-ARPES) can perform analogous measurements for
topological metals and insulators. We measured all of the
topological numbers for Bi$_{1-x}$Sb$_x$ and provide an
identification of its spin-texture, which heretofore was unmeasured
despite its surface states having been observed [10]. The measured
spin texture reveals the existence of a non-zero geometrical quantum
phase (Berry's phase [16,17]) and the handedness or chiral
properties. More importantly, this technique enables us to
investigate aspects of the metallic regime of the Bi$_{1-x}$Sb$_x$
series, such as spin properties in pure Sb, which are necessary to
determine the microscopic origin of topological order. Our
measurements on pure metallic Sb show that its surface carries a
geometrical (Berry's) phase and chirality property unlike the
conventional spin-orbit metals such as gold (Au), which has zero net
Berry's phase and no net chirality [18].

Strong topological materials are distinguished from ordinary
materials such as gold by a topological quantum number, $\nu_0$ = 1
or 0 respectively [14,15]. For Bi$_{1-x}$Sb$_x$, theory has shown
that $\nu_0$ is determined solely by the character of the bulk
electronic wave functions at the $L$ point in the three-dimensional
(3D) Brillouin zone (BZ). When the lowest energy conduction band
state is composed of an antisymmetric combination of atomic $p$-type
orbitals ($L_a$) and the highest energy valence band state is
composed of a symmetric combination ($L_s$), then $\nu_0$ = 1, and
vice versa for $\nu_0$ = 0 [11]. Although the bonding nature
(parity) of the states at $L$ is not revealed in a measurement of
the bulk band structure, the value of $\nu_0$ can be determined from
the spin-textures of the surface bands that form when the bulk is
terminated. In particular, a $\nu_0$ = 1 topology requires the
terminated surface to have a Fermi surface (FS) [1] that supports a
non-zero Berry's phase (odd as opposed to even multiple of $\pi$),
which is not realizable in an ordinary spin-orbit material.

In a general inversion symmetric spin-orbit insulator, the bulk
states are spin degenerate because of a combination of space
inversion symmetry $[E(\vec{k},\uparrow) = E(-\vec{k},\uparrow)]$
and time reversal symmetry $[E(\vec{k},\uparrow) =
E(-\vec{k},\downarrow)]$. Because space inversion symmetry is broken
at the terminated surface, the spin degeneracy of surface bands can
be lifted by the spin-orbit interaction [19-21]. However, according
to Kramers theorem [16], they must remain spin degenerate at four
special time reversal invariant momenta ($\vec{k}_T$ =
$\bar{\Gamma}$, \={M}) in the surface BZ [11], which for the (111)
surface of Bi$_{1-x}$Sb$_x$ are located at $\bar{\Gamma}$ and three
equivalent \={M} points [see Fig.1(A)].

Depending on whether $\nu_0$ equals 0 or 1, the Fermi surface
pockets formed by the surface bands will enclose the four
$\vec{k}_T$ an even or odd number of times respectively. If a Fermi
surface pocket does not enclose $\vec{k}_T$ (= $\bar{\Gamma}$,
\={M}), it is irrelevant for the topology [11,20]. Because the wave
function of a single electron spin acquires a geometric phase factor
of $\pi$ [16] as it evolves by 360$^{\circ}$ in momentum space along
a Fermi contour enclosing a $\vec{k}_T$, an odd number of Fermi
pockets enclosing $\vec{k}_T$ in total implies a $\pi$ geometrical
(Berry's) phase [11]. In order to realize a $\pi$ Berry's phase the
surface bands must be spin-polarized and exhibit a partner switching
[11] dispersion behavior between a pair of $\vec{k}_T$. This means
that any pair of spin-polarized surface bands that are degenerate at
$\bar{\Gamma}$ must not re-connect at \={M}, or must separately
connect to the bulk valence and conduction band in between
$\bar{\Gamma}$ and \={M}. The partner switching behavior is realized
in Fig. 1(C) because the spin down band connects to and is
degenerate with different spin up bands at $\bar{\Gamma}$ and \={M}.
The partner switching behavior is realized in Fig. 2(A) because the
spin up and spin down bands emerging from $\bar{\Gamma}$ separately
merge into the bulk valence and conduction bands respectively
between $\bar{\Gamma}$ and \={M}.

We first investigate the spin properties of the topological
insulator phase. Spin-integrated ARPES [19] intensity maps of the
(111) surface states of insulating Bi$_{1-x}$Sb$_x$ taken at the
Fermi level ($E_F$) [Figs 1(D)\&(E)] show that a hexagonal FS
encloses $\bar{\Gamma}$, while dumbbell shaped FS pockets that are
much weaker in intensity enclose \={M}. By examining the surface
band dispersion below the Fermi level [Fig.1(F)] it is clear that
the central hexagonal FS is formed by a single band (Fermi crossing
1) whereas the dumbbell shaped FSs are formed by the merger of two
bands (Fermi crossings 4 and 5) [10].

This band dispersion resembles the partner switching dispersion
behavior characteristic of topological insulators. To check this
scenario and determine the topological index $\nu_0$, we have
carried out spin-resolved photoemission spectroscopy. Fig.1(G) shows
a spin-resolved momentum distribution curve taken along the
$\bar{\Gamma}$-\={M} direction at a binding energy $E_B$ = $-$25 meV
[Fig.1(G)]. The data reveal a clear difference between the spin-up
and spin-down intensities of bands 1, 2 and 3, and show that bands 1
and 2 have opposite spin whereas bands 2 and 3 have the same spin
(detailed analysis discussed later in text). The former observation
confirms that bands 1 and 2 form a spin-orbit split pair, and the
latter observation suggests that bands 2 and 3 (as opposed to bands
1 and 3) are connected above the Fermi level and form one band. This
is further confirmed by directly imaging the bands through raising
the chemical potential via doping [see supporting online material
(SOM B) [22]]. Irrelevance of bands 2 and 3 to the topology is
consistent with the fact that the Fermi surface pocket they form
does not enclose any $\vec{k}_T$. Because of a dramatic intrinsic
weakening of signal intensity near crossings 4 and 5, and the small
energy and momentum splittings of bands 4 and 5 lying at the
resolution limit of modern spin-resolved ARPES spectrometers, no
conclusive spin information about these two bands can be drawn from
the methods employed in obtaining the data sets in Figs 1(G)\&(H).
However, whether bands 4 and 5 are both singly or doubly degenerate
does not change the fact that an odd number of spin-polarized FSs
enclose the $\vec{k}_T$, which provides evidence that
Bi$_{1-x}$Sb$_x$ has $\nu_0$ = 1 and that its surface supports a
non-trivial Berry's phase.

We investigated the quantum origin of topological order in this
class of materials. It has been theoretically speculated that the
novel topological order originates from the parities of the
electrons in pure Sb and not Bi [11,23]. It was also noted [20] that
the origin of the topological effects can only be tested by
measuring the spin-texture of the Sb surface, which has not been
measured. Based on quantum oscillation and magneto-optical studies,
the bulk band structure of Sb is known to evolve from that of
insulating Bi$_{1-x}$Sb$_x$ through the hole-like band at H rising
above $E_F$ and the electron-like band at $L$ sinking below $E_F$
[23]. The relative energy ordering of the $L_a$ and $L_s$ states in
Sb again determines whether the surface state pair emerging from
$\bar{\Gamma}$ switches partners [Fig.2(A)] or not [Fig.2(B)]
between $\bar{\Gamma}$ and \={M}, and in turn determines whether
they support a non-zero Berry's phase.

In a conventional spin-orbit metal such as gold, a free-electron
like surface state is split into two parabolic spin-polarized
sub-bands that are shifted in $\vec{k}$-space relative to each other
[18]. Two concentric spin-polarized Fermi surfaces are created, one
having an opposite sense of in-plane spin rotation from the other,
that enclose $\bar{\Gamma}$. Such a Fermi surface arrangement, like
the schematic shown in figure 2(B), does not support a non-zero
Berry's phase because the $\vec{k}_T$ are enclosed an even number of
times (2 for most known materials).

However, for Sb, this is not the case. Figure 2(C) shows a
spin-integrated ARPES intensity map of Sb(111) from $\bar{\Gamma}$
to \={M}. By performing a systematic incident photon energy
dependence study of such spectra, previously unavailable with He
lamp sources [24], it is possible to identify two V-shaped surface
states (SS) centered at $\bar{\Gamma}$, a bulk state located near
$k_x$ = $-$0.25 \AA$^{-1}$ and resonance states centered about $k_x$
= 0.25 \AA$^{-1}$ and \={M} that are hybrid states formed by surface
and bulk states [19] (SOM C [22]). An examination of the ARPES
intensity map of the Sb(111) surface and resonance states at $E_F$
[Fig.2(E)] reveals that the central surface FS enclosing
$\bar{\Gamma}$ is formed by the inner V-shaped SS only. The outer
V-shaped SS on the other hand forms part of a tear-drop shaped FS
that does \textit{not} enclose $\bar{\Gamma}$, unlike the case in
gold. This tear-drop shaped FS is formed partly by the outer
V-shaped SS and partly by the hole-like resonance state. The
electron-like resonance state FS enclosing \={M} does not affect the
determination of $\nu_0$ because it must be doubly spin degenerate
(SOM D [22]). Such a FS geometry [Fig.2(G)] suggests that the
V-shaped SS pair may undergo a partner switching behavior expected
in Fig.2(A). This behavior is most clearly seen in a cut taken along
the $\bar{\Gamma}$-\={K} direction since the top of the bulk valence
band is well below $E_F$ [Fig.2(F)] showing only the inner V-shaped
SS crossing $E_F$ while the outer V-shaped SS bends back towards the
bulk valence band near $k_x$ = 0.1 \AA$^{-1}$ before reaching $E_F$.
The additional support for this band dispersion behavior comes from
tight binding surface calculations on Sb [Fig.2(D)], which closely
match with experimental data below $E_F$. Our observation of a
single surface band forming a FS enclosing $\bar{\Gamma}$ suggests
that pure Sb is likely described by $\nu_0$ = 1, and that its
surface may support a Berry's phase.

Confirmation of a surface $\pi$ Berry's phase rests critically on a
measurement of the relative spin orientations (up or down) of the SS
bands near $\bar{\Gamma}$ so that the partner switching is indeed
realized, which cannot be done without spin resolution. Spin
resolution was achieved using a Mott polarimeter that measures two
orthogonal spin components of a photoemitted electron [27,28]. These
two components are along the $y'$ and $z'$ directions of the Mott
coordinate frame, which lie predominantly in and out of the sample
(111) plane respectively. Each of these two directions represents a
normal to a scattering plane defined by the photoelectron incidence
direction on a gold foil and two electron detectors mounted on
either side (left and right) [Fig.3(A)]. Strong spin-orbit coupling
of atomic gold is known to create an asymmetry in the scattering of
a photoelectron off the gold foil that depends on its spin component
normal to the scattering plane [28]. This leads to an asymmetry
between the left intensity ($I^L_{y',z'}$) and right intensity
($I^R_{y',z'}$) given by $A_{y',z'} =
(I^L_{y',z'}-I^R_{y',z'})/(I^L_{y',z'}+I^R_{y',z'})$, which is
related to the spin polarization $P_{y',z'} = (1/S_{eff})\times
A_{y',z'}$ through the Sherman function $S_{eff}$ = 0.085 [27,28].
Spin-resolved momentum distribution curve data sets of the SS bands
along the $-$\={M}-$\bar{\Gamma}$-\={M} cut at $E_B$ = $-$30 meV
[Fig.3(B)] are shown for maximal intensity. Figure 3(D) displays
both $y'$ and $z'$ polarization components along this cut, showing
clear evidence that the bands are spin polarized, with spins
pointing largely in the (111) plane. In order to estimate the full
3D spin polarization vectors from a two component measurement (which
is not required to prove the partner switching or the Berry's
phase), we fit a model polarization curve to our data following the
recent demonstration in Ref-[26], which takes the polarization
directions associated with each momentum distribution curve peak
[Fig.3(C)] as input parameters, with the constraint that each
polarization vector has length one (in angular momentum units of
$\hbar$/2). Our fitted polarization vectors are displayed in the
sample ($x,y,z$) coordinate frame [Fig.3(F)], from which we derive
the spin-resolved momentum distribution curves for the spin
components parallel ($I_y^{\uparrow}$) and anti-parallel
($I_y^{\downarrow}$) to the $y$ direction (SOM B [22]) as shown in
figure 3(E). There is a clear difference in $I_y^{\uparrow}$ and
$I_y^{\downarrow}$ at each of the four momentum distribution curve
peaks indicating that the surface state bands are spin polarized
[Fig.3(E)], which is possible to conclude even without a full 3D
fitting. Each of the pairs $l2/l1$ and $r1/r2$ have opposite spin,
consistent with the behavior of a spin split pair, and the spin
polarization of these bands are reversed on either side of
$\bar{\Gamma}$ in accordance with the system being time reversal
symmetric $[E(\vec{k},\uparrow) = E(-\vec{k},\downarrow)]$
[Fig.3(F)]. The measured spin texture of the Sb(111) surface states
(Fig.3), together with the connectivity of the surface bands
(Fig.2), uniquely determines its belonging to the $\nu_0$ = 1 class.
Therefore the surface of Sb carries a non-zero ($\pi$) Berry's phase
via the inner V-shaped band and pure Sb can be regarded as the
parent metal of the Bi$_{1-x}$Sb$_x$ topological insulator class, in
other words, the topological order originates from the Sb wave
functions.

Our spin polarized measurement methods (Fig.1 and 3) uncover a new
type of topological quantum number $n_M$ which provides information
about the chirality properties. Topological band theory suggests
that the bulk electronic states in the mirror ($k_y$ = 0) plane can
be classified in terms of a number $n_M$ (=$\pm$1) that describes
the handedness (either left or right handed) or chirality of the
surface spins which can be directly measured or seen in
spin-resolved experiments [20]. We now determine the value of $n_M$
from our data. From figure 1, it is seen that a single (one) surface
band, which switches partners at \={M}, connects the bulk valence
and conduction bands, so $|n_M|$ = 1 (SOM F [22]). The sign of $n_M$
is related to the direction of the spin polarization $\langle
\vec{P} \rangle$ of this band [20], which is constrained by mirror
symmetry to point along $\pm\hat{y}$. Since the central
electron-like FS enclosing $\bar{\Gamma}$ intersects six mirror
invariant points [see Fig.3(B)], the sign of $n_M$ distinguishes two
distinct types of handedness for this spin polarized FS. Figures
1(F) and 3 show that for both Bi$_{1-x}$Sb$_x$ and Sb, the surface
band that forms this electron pocket has $\langle \vec{P} \rangle
\propto -\hat{y}$ along the $k_x$ direction, suggesting a
left-handed rotation sense for the spins around this central FS thus
$n_M$ = $-$1. Therefore, both insulating Bi$_{1-x}$Sb$_x$ and pure
Sb possess equivalent chirality properties $-$ a definite spin
rotation sense (left-handed chirality, see Fig.3(B)) and a
topological Berry's phase.

These spin-resolved experimental measurements reveal an intimate and
straightforward connection between the topological quantum numbers ($\nu_0$,
$n_M$) and the physical observables. The $\nu_0$ determines whether
the surface electrons support a non-trivial Berry's phase, and if
they do, the $n_M$ determines the spin handedness of the Fermi
surface that manifests this Berry's phase. The 2D Berry's phase is a
critical signature of topological order and is not realizable in
isolated 2D electron systems, nor on the surfaces of conventional
spin-orbit or exchange coupled magnetic materials. A non-zero
Berry's phase is known, theoretically, to protect an electron system
against the almost universal weak-localization behavior in their low
temperature transport [11,13] and is expected to form the key
element for fault-tolerant computation schemes [13,29, 30]. Its remarkable realization on the Bi$_{1-x}$Sb$_x$ surface represents an unprecedented example of a 2D
$\pi$ Berry's phase, and opens the possibility for building
realistic prototype systems to test quantum computing modules.

\textbf{In summary, using Mott polarimetry, for the first time we probed the spin degrees of freedom of these quantum spin Hall states and demonstrated that topological quantum numbers are uniquely determined from spin texture imaging measurements. Applying this method to the Bi$_{1-x}$Sb$_x$ series, we identified the origin of its novel order and
unusual chiral properties. These results taken together constitute the first observation of surface electrons collectively carrying a geometrical quantum Berry's phase and definite chirality (topological mirror Chern number $n_M$ =-1), which are the key electronic properties for realizing topological computing bits with intrinsic spin Hall-like topological phenomena. Our spin-resolved results not only provides the first clear proof of a topological insulating state in nature but also demonstrate the utility of spin-resolved ARPES technique in measuring the quantum spin Hall phases of matter.
Our results as demonstrated here thus open up a new search front for topological quantum materials for novel spin-devices and paves a new way to achieving fault-tolerant quantum computing.}

\vspace{1cm}

We thank J. Teo for providing the SS band calculations of antimony; A. Fedorov, L. Patthey, and D.-H. Lu for beamline assistance; and D. Haldane, B. I. Halperin, N. P. Ong, D. A. Huse, F. Wilczek, P. W. Anderson, D. C. Tsui, J. E. Moore, L. Fu, L. Balents, D.-H. Lee, S. Sachdev, P. A. Lee, and X.-G. Wen for stimulating discussions. C.L.K. was supported by NSF grant DMR-0605066. The spin-resolved ARPES experiments are supported by NSF through the Center for Complex Materials (DMR-0819860/MZH) and Princeton University; the use of synchrotron facilities (ALS-LBNL, Berkeley, and SSRL-SLAC, Stanford) is supported by the Basic Energy Sciences of the U.S. Department of Energy (DE-FG-02–05ER46200/MZH) and by the Swiss Light Source, Paul Scherrer Institute, Switzerland.

\vspace{1cm}


\newpage

\textbf{FIG. 1.  Theoretical spin spectrum of a topological
insulator and spin-resolved spectroscopy results.} (A) Schematic
sketches of the bulk Brillouin zone (BZ) and (111) surface BZ of the
Bi$_{1-x}$Sb$_x$ crystal series. The high symmetry points
(L,H,T,$\Gamma$,$\bar{\Gamma}$,\={M},\={K}) are identified. (B)
Schematic of Fermi surface pockets formed by the surface states (SS)
of a topological insulator that carries a Berry's phase. (C) Partner
switching band structure topology: Schematic of spin-polarized SS
dispersion and connectivity between $\bar{\Gamma}$ and \={M}
required to realize the FS pockets shown in panel-(B). $L_a$ and
$L_s$ label bulk states at $L$ that are antisymmetric and symmetric
respectively under a parity transformation (see text). (D)
Spin-integrated ARPES intensity map of the SS of
Bi$_{0.91}$Sb$_{0.09}$ at $E_F$. Arrows point in the measured
direction of the spin. (E) High resolution ARPES intensity map of
the SS at $E_F$ that enclose the \={M}$_1$ and \={M}$_2$ points.
Corresponding band dispersion (second derivative images) are shown
below. The left right asymmetry of the band dispersions are due to
the slight offset of the alignment from the
$\bar{\Gamma}$-\={M}$_1$(\={M}$_2$) direction. (F) Surface band
dispersion image along the $\bar{\Gamma}$-\={M} direction showing
five Fermi level crossings. The intensity of bands 4,5 is scaled up
for clarity (the dashed white lines are guides to the eye). The
schematic projection of the bulk valence and conduction bands are
shown in shaded blue and purple areas. (G) Spin-resolved momentum
distribution curves presented at $E_B$ = $-$25 meV showing single
spin degeneracy of bands at 1, 2 and 3. Spin up and down correspond
to spin pointing along the +$\hat{y}$ and -$\hat{y}$ direction
respectively. (H) Schematic of the spin-polarized surface FS
observed in our experiments. It is consistent with a $\nu_0$ = 1
topology (compare (B) and (H)).

\vspace{1cm}

\textbf{FIG. 2.  Topological character of pure Sb revealed on the
(111) surface states.} Schematic of the bulk band structure (shaded
areas) and surface band structure (red and blue lines) of Sb near
$E_F$ for a (A) topologically non-trivial and (B) topological
trivial (gold-like) case, together with their corresponding surface
Fermi surfaces are shown. (C) Spin-integrated ARPES spectrum of
Sb(111) along the $\bar{\Gamma}$-\={M} direction. The surface states
are denoted by SS, bulk states by BS, and the hole-like resonance
states and electron-like resonance states by h RS and e$^-$ RS
respectively. (D) Calculated surface state band structure of Sb(111)
based on the methods in [20,25]. The continuum bulk energy bands are
represented with pink shaded regions, and the lines show the
discrete bands of a 100 layer slab. The red and blue single bands,
denoted $\Sigma_1$ and $\Sigma_2$, are the surface states bands with
spin polarization $\langle \vec{P} \rangle \propto +\hat{y}$ and
$\langle \vec{P} \rangle \propto -\hat{y}$ respectively. (E) ARPES
intensity map of Sb(111) at $E_F$ in the $k_x$-$k_y$ plane. The only
one FS encircling $\bar{\Gamma}$ seen in the data is formed by the
inner V-shaped SS band seen in panel-(C) and (F). The outer V-shaped
band bends back towards the bulk band best seen in data in
panel-(F). (F) ARPES spectrum of Sb(111) along the
$\bar{\Gamma}$-\={K} direction shows that the outer V-shaped SS band
merges with the bulk band. (G) Schematic of the surface FS of
Sb(111) showing the pockets formed by the surface states (unfilled)
and the resonant states (blue and purple). The purely surface state
Fermi pocket encloses only one Kramers degenerate point
($\vec{k}_T$), namely, $\bar{\Gamma}$(=$\vec{k}_T$), therefore
consistent with the $\nu_0$ = 1 topological classification of Sb
which is different from Au (compare (B) and (G)). As discussed in
the text, the hRS and e$^-$RS count trivially.

\vspace{1cm}

\textbf{FIG. 3.  Spin-texture of topological surface states and
chirality.} (A) Experimental geometry of the spin-resolved ARPES
study. At normal emission ($\theta$ = 0$^{\circ}$), the sensitive
$y'$-axis of the Mott detector is rotated by 45$^{\circ}$ from the
sample $\bar{\Gamma}$ to $-$\={M} ($\parallel -\hat{x}$) direction,
and the sensitive $z'$-axis of the Mott detector is parallel to the
sample normal ($\parallel \hat{z}$). (B) Spin-integrated ARPES
spectrum of Sb(111) along the $-$\={M}-$\bar{\Gamma}$-\={M}
direction. The momentum splitting between the band minima is
indicated by the black bar and is approximately 0.03 \AA$^{-1}$. A
schematic of the spin chirality of the central FS based on the
spin-resolved ARPES results is shown on the right. (C) Momentum
distribution curve of the spin averaged spectrum at $E_B$ = $-$30
meV (shown in (B) by white line), together with the Lorentzian peaks
of the fit. (D) Measured spin polarization curves (symbols) for the
detector $y'$ and $z'$ components together with the fitted lines
using the two-step fitting routine [26]. (E) Spin-resolved spectra
for the sample $y$ component based on the fitted spin polarization
curves shown in (D). Up (down) triangles represent a spin direction
along the +(-)$\hat{y}$ direction. (F) The in-plane and out-of-plane
spin polarization components in the sample coordinate frame obtained
from the spin polarization fit. Overall spin-resolved data and the
fact that the surface band that forms the central electron pocket
has $\langle \vec{P} \rangle \propto -\hat{y}$ along the +$k_x$
direction, as in (E), suggest a left-handed chirality (schematic in
(B) and see text for details).

\newpage

\begin{figure*}
\includegraphics[scale=0.58,clip=true, viewport=0.0in 0.5in 11.3in 6.5in]{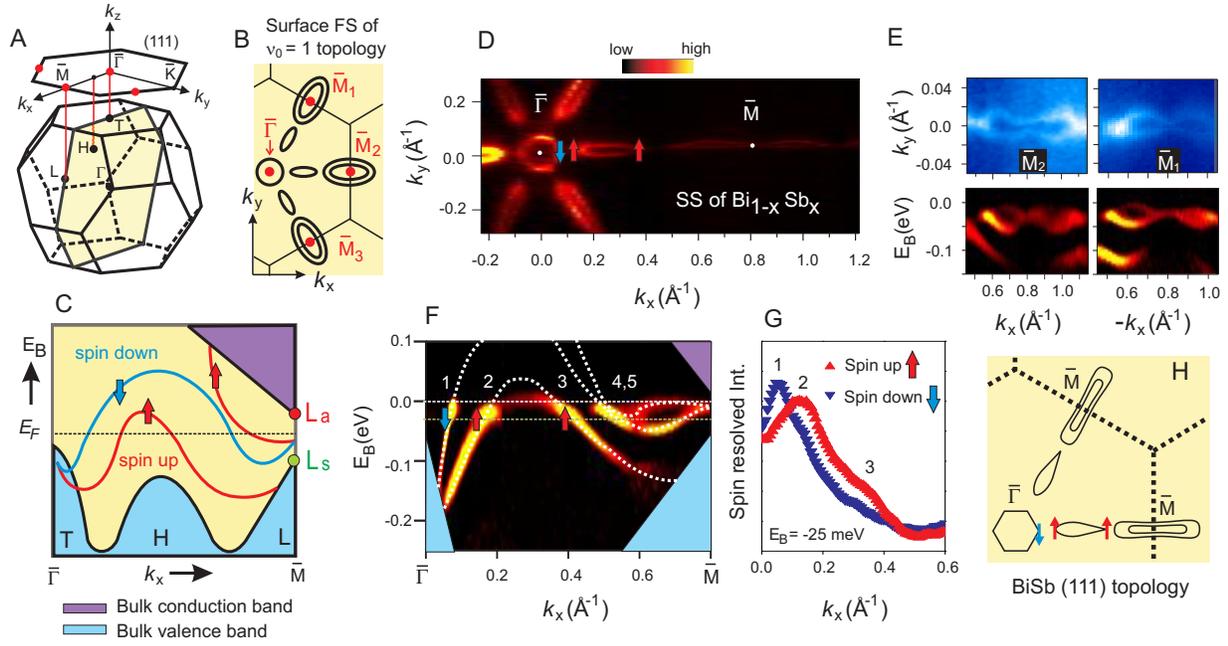}
\caption{D. Hsieh $et$ $al$. SCIENCE 323, 919 (2009) [Princeton University]}
\end{figure*}

\clearpage

\begin{figure*}
\includegraphics[scale=0.6,clip=true, viewport=0.0in 0.7in 11.0in 5.4in]{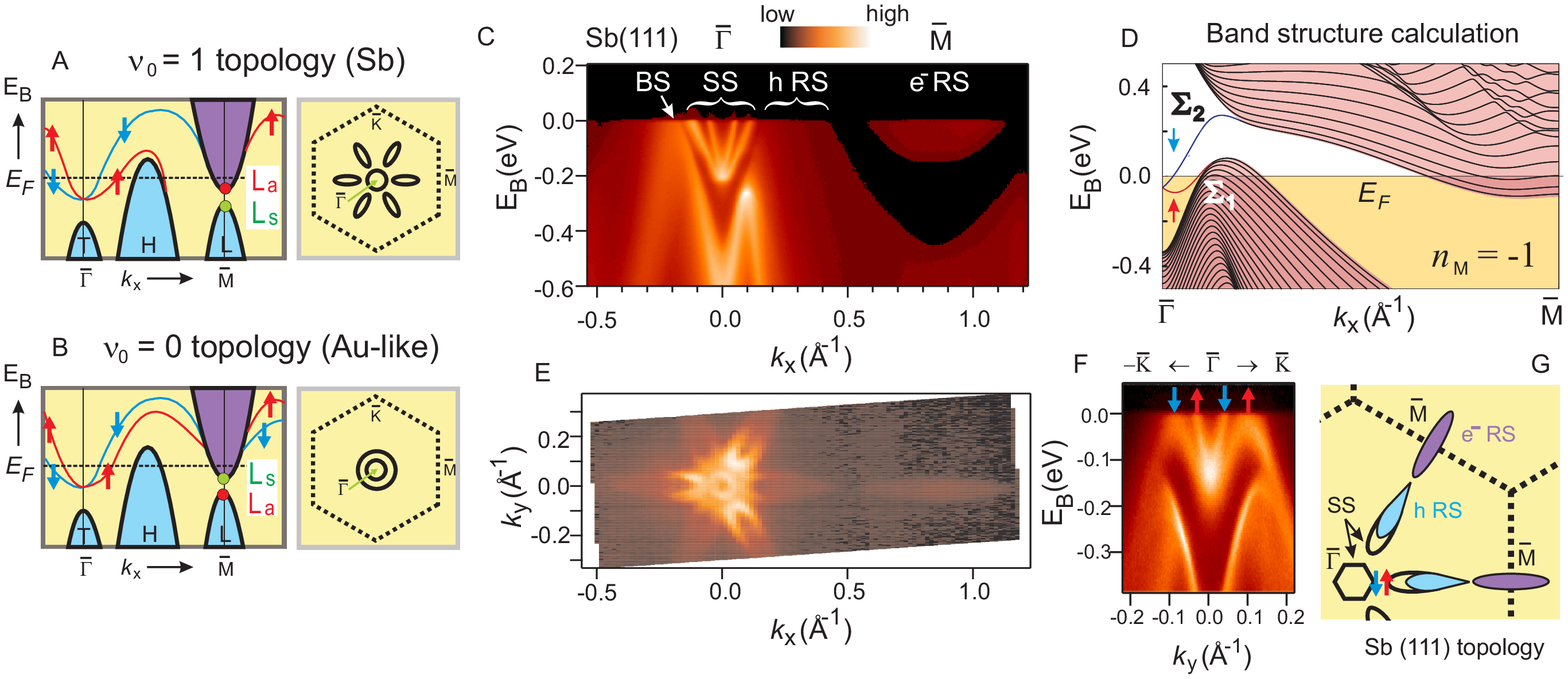}
\caption{D. Hsieh $et$ $al$. SCIENCE 323, 919 (2009) [Princeton University]}
\end{figure*}

\clearpage

\begin{figure*}
\includegraphics[scale=0.62,clip=true, viewport=0.0in 0in 11.0in 6.0in]{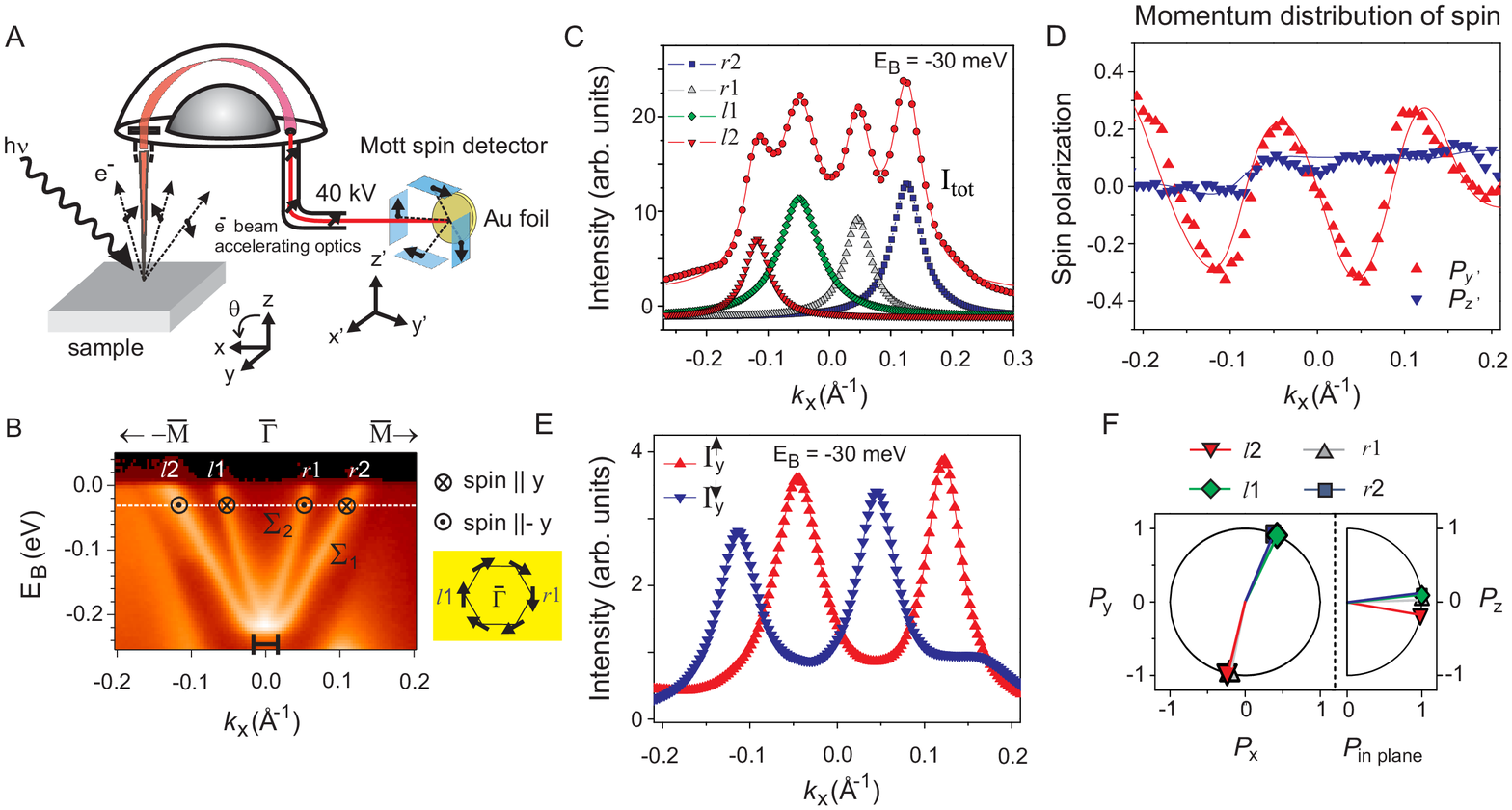}
\caption{D. Hsieh $et$ $al$. SCIENCE 323, 919 (2009) [Princeton University]}
\end{figure*}

\end{document}